\documentclass[a4paper]{article}

\usepackage{INTERSPEECH2020}
\usepackage[T1]{fontenc}
\usepackage[utf8]{inputenc}
\usepackage[english]{babel}
\usepackage{amsmath}
\usepackage{amsfonts}
\usepackage{amssymb}
\usepackage{cite}
\usepackage{textgreek}
\usepackage{verbatim}
\usepackage{graphicx}
\usepackage{multirow}
\usepackage{subscript}
\usepackage{url}
\usepackage{color}
\usepackage{mdframed}
\usepackage{makecell}

\title{	Speech Sentiment and Customer Satisfaction Estimation \\in Socialbot Conversations}
\name{Yelin Kim, Joshua Levy, Yang Liu}
\address{
 Amazon, Inc. }
\email{{kimyelin,levyjos,yangliud}@amazon.com}

\begin{document}

\maketitle
\begin{abstract}
For an interactive agent, such as task-oriented spoken dialog systems or chatbots, measuring and adapting to Customer Satisfaction (CSAT) is critical in order to understand user perception of an agent’s behavior and increase user engagement and retention. However, an agent often relies on explicit customer feedback for measuring CSAT. Such explicit feedback may result in potential distraction to users and it can be challenging to capture continuously changing user's satisfaction. To address this challenge, we present a new approach to automatically estimate  CSAT using acoustic and lexical information in the Alexa Prize Socialbot data. We first explore the relationship between CSAT and sentiment scores at both the utterance and conversation level. We then investigate static and temporal modeling methods that use estimated sentiment scores as a mid-level representation. The results show that the sentiment scores, particularly valence and satisfaction, are correlated with CSAT. We also demonstrate that our proposed temporal modeling approach for estimating CSAT achieves competitive performance, relative to static baselines as well as human performance. This work provides insights into open domain social conversations between real users and socialbots, and the use of both acoustic and lexical information for understanding the relationship between CSAT and sentiment scores. 
\end{abstract}

\section{Introduction}

For an interactive agent, such as task-oriented spoken dialog systems or chit chat bots, measuring and adapting to Customer Satisfaction (CSAT) is critical in order to understand and enhance user experience.
However, an agent often relies on explicit customer feedback for measuring CSAT. Such explicit feedback requires user actions and there are maximum frequency caps about how many times we can ask for such feedback, since frequent surveys often increase user annoyance. Further, survey questions can be ambiguous and may have inadequate response options, particularly for early-stage products. Also, customer surveys cannot be used in real time for capturing user’s immediate responses at the moment.
Therefore automatic estimation of CSAT is beneficial to improve system behavior.

In this study, we investigate if user's sentiment is correlated with their utterance-level or conversation-level satisfaction using socialbot conversations, and present an automatic estimation model for CSAT  in such conversations. 
Intuitively positive sentiments captured naturally during the interactions can provide insights into delightful features of the agent, whereas negative sentiments may indicate defect or dissatisfaction.
A lot of previous work on CSAT prediction has been conducted using call center customer service data \cite{Zweig06naacl,Park07cikm,Ando2017interspeech}, where the users typically have a clear goal, and thus there is less ambiguity about users' satisfaction. 
In contrast, in socialbot conversations, the goal or task of the conversation is not well defined, and thus open questions remain about the relationship of the utterance-level user sentiment and conversation-level CSAT.  
For chat bot applications, previous studies recognized the importance of sentiment factors and developed systems that can recognize user's sentiment and incorporate the recognized sentiment to generate system responses \cite{zhou2018emotional,asghar2018affective,Sankar2019sigdial,Feine2019,FBblender}.  These studies demonstrated that users rated higher for the systems with the sentiment-aware skills. 

To address the open questions in socialbot applications, we aim to test two hypotheses in this work:  
\textbf{(H1)} user’s sentiment  is correlated with CSAT 
and \textbf{(H2)}  conversation-level CSAT can be automatically predicted from user’s sentiment-- both from the ground truth (annotated, perceived) sentiment and the automatically estimated sentiment from acoustic and lexical cues.  

We use naturally occurring human-machine conversations collected from Alexa Prize (AP) socialbots \cite{alexaprize, venkatesh2018evaluating, yi2019towards}. These conversations have CSAT scores reported by a user in response to a feedback request from Alexa. 
We conduct experiments on two conditions, one with sentiment annotation and another without sentiment annotation. 
For the latter case, we use a two-step processing of (i) generating sentiment embeddings of each utterance using lexical and audio information based on Long Short Term Memory networks (LSTMs) and (ii) using the estimated sentiment embeddings as a mid-level representation to infer the overall CSAT at the conversation level using both static (\textnu-Support Vector Regression, or \textnu-SVR) and temporal (Bi-direction LSTM, or BLSTM) regression models.

Our results demonstrate that valence and satisfaction scores are correlated with CSAT, with a correlation of up to 0.2847 for annotated satisfaction scores. 
Additionally,  our proposed regression model using BLSTM outperforms the static \textnu-SVR model by 12.56\%, and is even slightly better than that from human annotators. 
The correlation between the self-reported user CSAT and the annotators' estimated CSAT is 0.2244, indicating how challenging it is to infer CSAT in socialbot conversations.   
To the best of our knowledge, this is the first work that demonstrates how CSAT can be automatically  predicted using sentiment embeddings based on both acoustic and lexical cues in socialbot conversations. 
Such sentiment recognition models are unintrusive and continuous in contrast to explicit customer feedback,  and allow customers to naturally address defects and dissatisfaction by themselves without any actions, immediately and in a scalable way.
This study also highlights  differences and challenges between socialbot conversations and other domains.

\section{ Background}

\subsection{SocialBot in Alexa Prize}
\label{sec:alexaprize}
Amazon launched a yearly competition called Alexa Prize in 2016. The grand challenge objective is to build agents (socialbots) that can converse coherently and engagingly with humans for 20 minutes, and obtain a 4 out of 5 or higher rating from humans interacting with them. Alexa customers interact with socialbots with invocation phrases ``Alexa, let’s chat'' (or similar ones). They are read a brief message about the competition (the introductory message varies depending on the completion phases), and instructions on how to end the conversation and provide ratings and feedback. Then they are connected to one of the participating socialbots. Customers may end the conversation at any time, and afterwards, the user is prompted to provide a verbal rating from 1 to 5: ``How do you feel about speaking with this socialbot again?'', followed by an option to provide additional freeform feedback.  Such ratings and feedback have been given to the participating teams to help them improve their socialbots. Note these ratings and feedback are optional, and only some users provide these.  
See \cite{alexaprize} for more information on the competition and the participating systems.  
In this study, we use the customer’s rating at the end of these conversations as an approximation for CSAT. 

\vspace{-0.2cm}
\subsection{Related Work}

Given the importance of understanding user's satisfaction of a service, previous studies explored methods to automatically predict CSAT~ \cite{Zweig06naacl,Park07cikm,Ando2017interspeech}. 
Most of previous work focuses on  call center applications, where the conversations are between customers and agents. 
To predict user's overall CSAT with the call, previous work treated this as a classification or regression task, i.e., providing a label or score for the entire dialog. They used various features including textual information derived from the transcriptions, and speech and prosodic cues from the audio, and explored different classification models.
In addition to conversation-level satisfaction, Ando et al.~performed utterance level and dialog level classification jointly~\cite{Ando2017interspeech}.   
Recently similar work has been conducted using task oriented chat bots with text-based interaction~\cite{Feine2019}.
Our work is different from these in that our application is on open domain conversations in a socialbot setting. 
The main difference of our work from previous studies on customer service calls or chats is that previous studies focus on  users who have a task or issues to resolve and thus satisfaction is more clearly defined, whereas for socialbots, the notion of dialog-level satisfaction is not clear, since users typically do not have goals and may chat about multiple topics. 
Recent work also pointed out the potential noise of user ratings and the low agreement between user's ratings and external evaluator's perception ~\cite{finch2020towards,liang2020beyond}. 
Therefore, it is an open question whether we can predict the overall satisfaction, and if utterance-based sentiment is correlated with the CSAT.  We will explore these questions in this study using real-world Socialbot data.

Our work also uses the estimated utterance-level sentiment  as a mid-level representation for predicting CSAT. Speech sentiment recognition is an active research field, and has  recently gained lots of advancements in recognition performance, particularly with the use of deep learning approaches~\cite{han2014speech, le2017discretized, huang2017deep}. 
Recent work focuses more on spontaneous data, particularly natural speech dialogs~\cite{lotfian2017building,batliner2008releasing, mckeown2011semaine,ringeval2013introducing}.  
Our approach contrasts with most
of the previous speech sentiment recognition work that has primarily focused on utterance-level recognition of emotion. Instead, we focus on the understanding of the conversation-level CSAT, and the relationship between utterance-level sentiment and CSAT.

Another line of work that is related to our high level motivation, but not this specific study, is developing systems that respond to a user's emotion properly.   
The early adopted voice interactive systems typically detected user's anger or frustration and took actions, for example, routing a call to human representatives. 
Recently there is more work in automatic response generation in conversational systems where user's sentiment information has been used as an additional attribute to control the system response such that the system responds to users emotionally appropriate 
\cite{asghar2018affective,zhou2018emotional,Sankar2019sigdial}.  
The recent work in \cite{FBblender} shows empathy is an important skill for socialbots.   
Note that most of such work is for text-based bots.  
In this study, we use speech for user sentiment recognition in open domain socialbots.

  \vspace{-0.2cm}
\section{Data}

\subsection{Alexa Prize (AP) Socialbot Data}
\label{sec:APdata}

We randomly selected some AP conversations with user ratings for this study. 
The data includes 6,308 AP conversations and 93,671 utterances, corresponding to an average conversation length of 14.85 utterances.  
This dataset is not annotated for ground truth sentiment, and hence we use sentiment scores automatically computed using our model described in Section \ref{sec:method:sentimentreco}.
We call this dataset as ``{unannotated} AP data'' in this paper.

We also use a smaller AP dataset with human annotated sentiment scores. This annotated dataset, which we call ``{annotated} AP data'', consists of 952 AP conversations and 14,415 utterances, corresponding to an average conversation length of 15.41 utterances.  
For both sets of data, we exclude conversations less than 5 utterances or greater than 50 utterances. This allows us to focus on conversations that have meaningful turns between users and socialbots and capture the overall conversation-level CSAT.   Additionally, we remove the two types of feedback utterances (Section \ref{sec:alexaprize}) from these datasets. This ensures that our prediction models do not have any information about the ground truth CSAT.

  \vspace{-0.2cm}
\subsection{Sentiment Annotation}
\label{sec:annotation}
For annotated AP data, our annotators rate each utterance of the conversation by considering both the tone and lexical content in the audio clip.  Each utterance was rated by at least three annotators on three dimensions, activation, valence, and satisfaction, on a $-3$ to $+3$ scale (where $0$ is neutral). Inter-annotator agreement is reported in Table \ref{tab:sentmodel}. 

Activation (excited vs. calm) and valence (positive vs. negative) are commonly used in the speech emotion recognition literature \cite{wu2010acoustic,ghosh2016representation,eyben2010line}. 
We add satisfaction as another sentiment dimension. Satisfaction is similar to valence, but is  designed  specifically for a voice assistant setting. We instruct the annotators that satisfaction is ``an attribute of speech that provides insight into the speaker’s level of satisfaction/pleasure (positive satisfaction) or dissatisfaction/displeasure (negative satisfaction) with Alexa for a speech segment.   Satisfaction can range from tones that convey frustration and disgust to tones that convey happiness and delight.'' Hence, satisfaction captures the degree of pleasure or displeasure with the voice assistant, whereas valence captures the user’s positiveness in general.  

  \vspace{-0.2cm}

\section{Methods}

\begin{figure}[t]
  \centering
  \includegraphics[width=\linewidth]{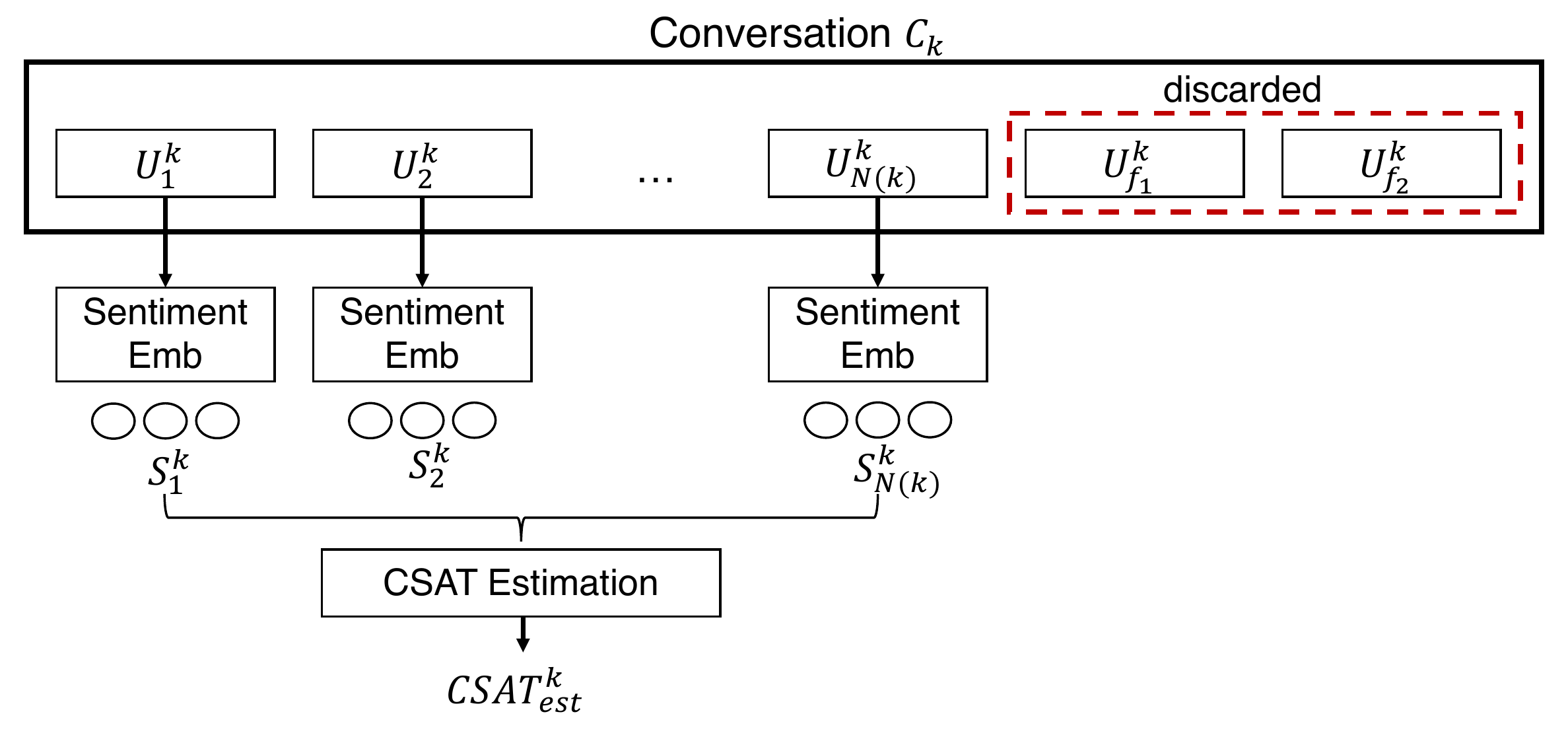}
  \vspace{-0.7cm}
  \caption{Overall two-step procedure for our CSAT estimation. For each conversation $C_k$, we first generate sentiment embeddings for each utterance and use these embeddings for final CSAT estimation at the conversation level. }
  \vspace{-0.3cm}
  \label{fig:overall}
\end{figure}

Figure \ref{fig:overall} shows our proposed two-phased modeling architecture to predict the overall CSAT from user's speech in the socialbot conversation. 
In the first phase, we apply the sentiment prediction model trained from  Alexa data (Section \ref{sec:method:sentimentreco}) to each user utterance in the conversation to obtain the sentiment scores.  
In the second phase, we use the sentiment embeddings for each utterance to estimate conversation-level CSAT (Section \ref{sec:method:csatpred}).

Let us denote a conversation $C^k$, a $k$-th conversation in Alexa Prize Socialbot data. 
$C^k$ is a sequence of utterances $U_t^k$ and $U_{f_1}^k$, $U_{f_2}^k$, where $t=1, 2, ,\dotsc,N(k)$. $N(k)$ denotes that the number of utterances varies for each conversation. 
$U_{f_1}^k$, $U_{f_2}^k$ are two feedback utterances described in Section \ref{sec:alexaprize} where the user responds to: ``On a scale from 1 to 5 stars, how do you feel about speaking with this SocialBot'' and ``do you have any feedback''. 
In our experiments,  we remove these feedback utterances in each conversation, 
and  compute the speech sentiment embeddings $S_t^k$ for each of $U_t^k$ (Section \ref{sec:method:sentimentreco}). 

  \vspace{-0.2cm}

\subsection{Speech Sentiment Embedding}
\label{sec:method:sentimentreco}

We train our sentiment model  using both acoustic and lexical cues at the utterance level.   On the acoustic side, we extract 40-dimensional Log Mel-filter bank energies (LFBEs) from the raw audio, and stack three frames of 10ms each to form an input that is 120 $\times$ (number of frames).  On the lexical side, we apply the 300-dimensional word2vec embeddings from Google news \cite{mikolov2013distributed} on Alexa’s Automatic Speech Recognition (ASR) output of the utterance.  
Long Short-Term Memory networks (LSTMs) are used to model these acoustic and lexical representations, with five layers of LSTMs for acoustic and one for lexical information.
Next, the outputs from the acoustic and lexical LSTMs are concatenated and passed through a separate dense layer for each sentiment type, activation, valence, and satisfaction respectively. Following \cite{lu2020speech}, we pre-trained our model on an ASR task and used the generated embeddings for the downstream sentiment recognition task.
The model was trained using 1k+ hours of general Alexa traffic annotated for sentiment, as well as some Alexa Prize data described in Section \ref{sec:APdata}.  

We use the average of human annotation as a ground truth of human annotation. 
We compare the model’s correlation with human annotation (``model agreement'') against the human-human agreement (``human agreement'') to measure the model’s performance.For both these metrics, we use Concordance Correlation Coefficient (CCC), which is similar to traditional Pearson's correlation but also factors in how well the mean and variance of the compared distributions match to each other \cite{lawrence1989concordance}.  The performance on the annotated AP data is shown in Table \ref{tab:sentmodel}. The sentiment model achieves competitive performance compared to human agreement, and is even better than human performance for activation.

\begin{table}[t]
\centering
 % \vspace{-0.2cm}
\caption{Sentiment recognition  performance using CCC metric. `HA' : human agreement and `MA': model agreement }
  \vspace{-0.2cm}
\begin{tabular}{c|ccc}
\hline
 & Activation & Valence & Satisfaction  \\ \hline
HA & 0.4891 & 0.5905 & 0.6009 \\
MA & 0.6562 & 0.5055 & 0.5406 \\ \hline
\end{tabular}
  \vspace{-0.5cm}
\label{tab:sentmodel}

\end{table}

%%%%%%%%%%%%%%%%%%%%%%%%%%%%%%%%%%%%%%%%%%%%%%
  \vspace{-0.3cm}
\subsection{Estimation of Customer Satisfaction (CSAT)}
\label{sec:method:csatpred}

For each conversation $C_k$, we use static and temporal regression models to predict an estimated $CSAT^k$ score. The regression models use the sentiment embeddings $S_t^k$ we compute from the model described in Section \ref{sec:method:sentimentreco}.

We use \textnu-Support Vector Regression (SVR) for static regression and explore performance of linear, polynomial, sigmoid and Radial Basis Function (RBF) kernels.  The parameter \textnu~ controls the fraction of training errors and support vectors. 

For temporal regression, we use the BLSTM model, which has been used widely in a variety of sequential models due to its ability to take into account both past and future information when learning the temporal structure~\cite{wollmer2012analyzing, zhang2018gender, brueckner2014social, lee2015high}.  
We experiment with both 2-dimensional (valence, satisfaction) and 3-dimensional (valence, activation, satisfaction) input features for BLSTMs. 
Since the input sequence lengths vary with high standard deviation (annotated AP data: 7.22, unannotated AP data: 18.18), we first sort the training data by sequence length so that a mini-batch has a similar length.
Within each mini-batch ($size = 10$), we pad the sequences with zero so that they have the same length.  
We use the Adam optimizer,  and clip the gradient to $threshold = 1$ if the gradient exceeds this. 
A maximum $30$ epochs are used for training. We use a mini-batch to evaluate the gradient of the loss function and update the weights.

%%%%%%%%%%%%%%%%%%%%%%%%%%%%%%%%%%%%%%%%%%%%%%

\vspace{-0.2cm}
\section{Results and Discussions}
\label{sec:results}

\subsection{Correlation Between Sentiment Scores and CSAT}
\label{ssec:corr}
\vspace{-0.2cm}

\begin{table}
\centering
\caption{Spearman's correlation between annotated sentiment scores (Act: Activation, Val: Valence, Sat: satisfaction) and Conversation-Level CSAT at the conversation and utterance levels  in {\bf Annotated} AP data .}
  \vspace{-0.2cm}

\begin{tabular}{c|ccc}
\hline
&	(Val, CSAT)	& (Act, CSAT) &	(Sat, CSAT) 	 \\ \hline
Conversation  &	\textbf{0.2584}	& 0.044 &	\textbf{0.2847} \\
Utterance &	0.0987 &	0.0236 &	0.1133 \\
 \hline
\end{tabular}
\vspace{-0.4cm}
\label{tab:corr_all_anno}
\end{table}

\begin{table}
\centering
%\vspace{-0.2cm}
\caption{Spearman's correlation between sentiment scores (Act: Activation, Val: Valence, Sat: satisfaction) and Conversation-Level CSAT at the conversation and utterance levels in {\bf Unannotated} AP data . }
  \vspace{-0.2cm}

\begin{tabular}{c|ccc}
\hline
&	(Val, CSAT)	& (Act, CSAT) &	(Sat, CSAT) 	 \\ \hline
Conversation &	\textbf{0.1919} & 0.046 & 0.1677 \\
Utterance &	0.0734 & 0.0427 & 0.0686\\
 \hline
\end{tabular}
 \vspace{-0.5cm}

\label{tab:corr_all_unanno}
\end{table}

In this section, we test  \textbf{(H1)} using both annotated and unannotated AP data. 
Tables \ref{tab:corr_all_anno} shows two correlations computed between the  conversation-level aggregated mean sentiment scores and CSAT, and the utterance-level sentiment scores and CSAT using the annotated data. 
Looking at the conversation-level results, satisfaction is most highly correlated to CSAT (correlation of 0.2846), the second highest is valence (correlation of 0.2584), 
 and activation scores are not correlated with CSAT. This may indicate that high activation can be derived from both  extreme joy and anger, and hence activation may not be a good indicator for measuring CSAT.  
 We can see that overall the utterance-level correlations are significantly lower than the ones at the conversation level, indicating that utterance-level scores and its characteristics change over time, and hence our prediction models should capture such variations throughout the overall conversation.

For unannotated AP data, Table  \ref{tab:corr_all_unanno} shows that the correlation is generally lower due to the machine prediction errors in our sentiment embeddings. 
Similar to Annotated AP data, the conversation-level correlations are higher than the ones at the utterance level. 
For the conversation level, valence is the most highly correlated sentiment to CSAT (correlation of 0.1919),  with satisfaction being the second (correlation of 0.1677). 
At the utterance level,  all of the predicted sentiment scores are not correlated with CSAT (correlation is less than 0.1 for valence, satisfaction, and activation). 

Overall, the results show that the aggregated (averaged) conversation-level emotion scores (valence and satisfaction) are correlated with CSAT ratings, both for annotated ground truth scores and predicted sentiment scores. 

  \vspace{-0.2cm}

\subsection{Prediction : Static and Temporal Regression Models}
\label{ssec:prediction}

In this section, we test \textbf{(H2)} that conversation-level CSAT can be automatically predicted from user’s sentiment. 
We use a regression-based static model using \textnu-SVR as our baseline method. We compare this with a BLSTM-based temporal model to understand the importance of capturing temporal structure in CSAT estimation.
Five-fold cross validation was used throughout the experiments for fair comparison. 
The averaged prediction performance measured using the Spearman's rank correlation coefficient across these five folds is used as metrics. 
Table \ref{tab:pred_all} summarizes the performance of our proposed  models.

To understand the human performance on estimating CSAT, we asked our annotators to   assess what they thought the customer satisfaction for the conversation would be.  We obtained these scores for 929 out of the 952 annotated conversations.  The Spearman’s correlation between the self-reported user CSAT score and the annotators' estimated CSAT score was 0.2244 ($p=4.56e^{-12}$).  This result is included in Table \ref{tab:pred_all}. It indicates that it is difficult for humans to judge users' CSAT scores. Note that users' CSAT is based on a variety of factors, such as the quality of responses or how engaging the conversation was, which may be difficult for third-party annotators to capture. This is also observed in recent studies on open domain socialbots  ~\cite{finch2020towards,liang2020beyond}.

\begin{table}[t]
\centering
\caption{Performance comparison between human performance and regression models using \textnu-SVR and BLSTM.} 
  \vspace{-0.2cm}

\begin{tabular}{c|ccc}
\hline
 & Human & \textnu-SVR  & BLSTM \\ \hline
 Annotated & 0.2244 & \textbf{0.2818} & 0.2759 \\
Unannotated & N/A & 0.2102 & \textbf{0.2366} \\
\hline
\end{tabular}
  \vspace{-0.5cm}

\label{tab:pred_all}
\end{table}

The results from the annotated AP data provide upper-bound performance of our CSAT prediction model, whereas the results from the unannotated AP data provide insights into how CSAT can be automatically estimated from acoustic and lexical cues.  
We use Spearman's correlation using the ranked scores of both sentiment and CSAT scores to capture the ordinal nature of these scores.

For annotated AP data, \textnu-SVR models achieve a slightly higher correlation (0.2818) compared to human correlation (0.2244), although direct comparison is not possible due to the difference in test data. 
BLSTM shows a slightly lower correlation (0.2759) compared to the best correlation from \textnu-SVR. 
This may indicate that  complex models like BLSTM cannot learn the relationship between the features and labels well given the limited data and feature size.  
As described in Section \ref{sec:APdata}, the annotated AP data is limited, with about 760 training instances for each fold.
We also empirically find that mean features of valence and satisfaction scores achieve higher correlation than feature sets with higher dimensionality, e.g., mean, standard deviation, quantile ranges of valence, satisfaction, and activation.  

On the other hand,  for the larger unannotated AP data,
 the BLSTM-based temporal regression model yielded 12.56\% relative improvement  compared to \textnu-SVR. 
We achieve the highest prediction performance of \textbf{0.2366} using valence, activation, and satisfaction scores as three-dimensional input features and 20 hidden units in BLSTM. 
Overall we observe more performance gain from using the temporal model in large unannotated AP compared to the static one, as well as the poor performance in small annotated data (or any model that is more complex than two-dimensional linear \textnu-SVR). This may  indicate that the temporal structure is important in this modeling of CSAT, and data size is important to capture that temporal structure.

Furthermore, we expect it is important to automatically identify the conversations that are poorly rated (e.g., $CSAT < 2$) or highly rated (e.g., $CSAT > 4$) in socialbot conversations or other user interactions. This can allow us to adapt our responses when such non-neutral CSAT scores are detected.
Table \ref{tab:pred_all_CS} shows our prediction power when CSAT scores are (1) {CSAT-R1}: $CSAT \leq 2$ or $CSAT \geq 3$ and (2) {CSAT-R2}: $CSAT < 2$ or $CSAT > 4$.
For annotated data, sigmoid-kernel \textnu-SVR performs the best, yielding a correlation of 0.3076 for CSAT-R1 and 0.3513 for CSAT-R2, which is higher than 0.2818 for all the data (9.16\% and 24.66\% increase in correlation, respectively). 
On the unannotated data, BLSTM-based model using predicted scores achieves a correlation of  0.2565 and 0.3088 on CSAT-R1 and CSAT-R2, respectively, which are both higher than \textnu-SVR-based correlation (0.2307 and 0.2880, respectively).

\begin{table}[t]
\centering
\caption{{Performance comparison between human performance and regression models using \textnu-SVR and BLSTM  for a subset of dataset that has (1) {CSAT-R1}: $CSAT \leq 2$ or $CSAT \geq 3$ and (2) {CSAT-R2}: $CSAT < 2$ or $CSAT > 4$.  } } 
\begin{tabular}{cc|ccc}
\hline
 &  & Human & \textnu-SVR & BLSTM \\
 \hline
Annotated & CSAT-R1 & 0.2418 & 0.3076 & 0.3031 \\
 & CSAT-R2 & 0.2711 & \textbf{0.3513} & \textbf{0.3501} \\
\hline
Unannotated & CSAT-R1 & N/A & 0.2307 & 0.2565 \\
 & CSAT-R2& N/A & 0.2880 & \textbf{0.3088} \\
 \hline
\end{tabular}
\vspace{-0.6cm}
\label{tab:pred_all_CS}
\end{table}

\vspace{-0.2cm}
\section{Conclusion and Future Work}
In this work, we empirically studied the correlation between turn-level sentiment and CSAT scores in real-world, open-domain socialbot conversations in Alexa Prize. 
We found that valence and satisfaction show the correlation of 0.2847 and 0.2584 with CSAT, respectively.
In addition, we evaluated regression models, static \textnu-SVR and temporal BLSTM, to predict CSAT based on sentiment embeddings provided by an automatic sentiment recognition system. 
Our experiments show that the BLSTM model trained on top of the sentiment embeddings outperforms the static models. This indicates that  the temporal structure of how sentiment changes over time within the conversation is important to estimate the overall  CSAT of the user. 
It is particularly challenging to infer customer satisfaction in socialbot settings, since the end goal of the conversations is open.  Consequently, we observed that it was difficult even for human annotators to predict customer CSAT with high correlation.  Our proposed method leveraged automatically generated sentiment embeddings to construct a model that predicted CSAT nearly as well as humans could.

Future work will explore whether CSAT prediction can be improved by integrating other factors besides the expressed sentiment, e.g. the quality of the responses provided by the socialbot.  Such factors could be included in addition to sentiment to enhance the ability to automatically predict customer satisfaction in conversational settings. 
Future work will also investigate what precise aspects of the temporal structure within the socialbot conversation are meaningful for understanding the CSAT.

\bibliographystyle{IEEEtran}

\bibliography{IS20paper_v4}

\end{document}